\documentclass[aps,prl,twocolumn,superscriptaddress]{revtex4-1}

\usepackage{natbib}

\pdfoutput=1
\usepackage{graphicx}
\usepackage{moresize}
\usepackage{amsmath}
\usepackage{bm}
\usepackage[caption=false]{subfig}
\usepackage{newtxtext}
\usepackage{newtxmath}

\usepackage{cancel}
\usepackage[usenames]{color}

\definecolor{bluegray}{rgb}{0.4, 0.6, 0.8}

\def\bea{\begin{eqnarray}}
\def\eea{\end{eqnarray}}
\def\ba{\begin{array}}
\def\ea{\end{array}}

\begin{document}

\title{Strain-controlled critical slowing down in the rheology of disordered networks}


\author{Jordan L.\ Shivers}
\altaffiliation[Present affiliation: ]{James Franck Institute and Department of Chemistry, University of Chicago, Chicago, Illinois 60637, USA}
\affiliation{Department of Chemical and Biomolecular Engineering, Rice University, Houston, TX 77005, USA}
\affiliation{Center for Theoretical Biological Physics, Rice University, Houston, TX 77005, USA}
\author{Abhinav Sharma}
\affiliation{Institute of Physics, University of Augsburg, 86159 Augsburg, Germany}
\affiliation{Leibniz-Institut für Polymerforschung Dresden, Institut Theorie der Polymere, 01069 Dresden, Germany}
\author{Fred C.\ MacKintosh}
\affiliation{Department of Chemical and Biomolecular Engineering, Rice University, Houston, TX 77005, USA}
\affiliation{Center for Theoretical Biological Physics, Rice University, Houston, TX 77005, USA}
\affiliation{Department of Chemistry, Rice University, Houston, TX 77005, USA} 
\affiliation{Department of Physics \& Astronomy, Rice University, Houston, TX 77005, USA}

\begin{abstract}

Networks and dense suspensions frequently reside near a boundary between soft (or fluid-like) and rigid (or solid-like) regimes. Transitions between these regimes can be driven by changes in structure, density, or applied stress or strain. In general, near the onset or loss of rigidity in these systems, dissipation-limiting heterogeneous nonaffine rearrangements dominate the macroscopic viscoelastic response, giving rise to diverging relaxation times and power-law rheology. Here, we describe a simple quantitative relationship between nonaffinity and the excess viscosity. We test this nonaffinity-viscosity relationship computationally and demonstrate its rheological consequences in simulations of strained filament networks and dense suspensions. We also predict critical signatures in the rheology of semiflexible and stiff biopolymer networks near the strain stiffening transition.

\end{abstract}

\maketitle

Polymer gels, suspensions, emulsions, and foams are inherently composite in nature, with both elastic and fluid-like components \cite{larson_structure_1999,chen_rheology_2010}. In these systems, minor variations in parameters such as volume fraction \cite{mooney_viscosity_1951,bagnold_experiments_1954,durian_foam_1995,paredes_rheology_2013}, connectivity \cite{maxwell_calculation_1864,thorpe_continuous_1983,alexander_amorphous_1998,wyart_elasticity_2008,broedersz_criticality_2011,zaccone_approximate_2011}, and applied strain \cite{biot_maurice_a_mechanics_1965,sharma_strain-controlled_2016,merkel_minimal-length_2019} can drive macroscopic transitions between fluid-like and solid-like behavior. These transitions are often heralded by familiar features of critical phenomena \cite{wilson_problems_1979,fisher_theory_1967,kadanoff_static_1967}, including power-law scaling of relevant quantities with distance to a critical point \cite{ohern_jamming_2003,drocco_multiscaling_2005,wyart_elasticity_2008,broedersz_criticality_2011,sharma_strain-controlled_2016} and diverging length and time scales \cite{hatano_growing_2009,olsson_relaxation_2015,bonn_yield_2017,vinutha_timescale_2019,nordstrom_microfluidic_2010,vinutha_timescale_2019,ikeda_universal_2020,saitoh_stress_2020}.  As a consequence of their disorder, these materials dissipate energy via heterogeneous or \textit{nonaffine} deformation, such that microscopic and macroscopic deformation fields differ \cite{alexander_amorphous_1998}. The associated microscopic nonaffine displacements can grow dramatically in magnitude near the onset or loss of rigidity and strongly influence macroscopic viscoelastic behavior \cite{tighe_dynamic_2012,yucht_dynamical_2013,milkus_atomic-scale_2017,palyulin_parameter-free_2018}. However, these displacements are neglected in continuum models and are notoriously difficult to measure in experiments \cite{wen_local_2007,liu_visualizing_2007,basu_nonaffine_2011} except in special cases, such as confocal microscopy of colloidal suspensions \cite{schall_structural_2007,chikkadi_long-range_2011,chikkadi_nonaffine_2012}.

Indirect evidence of nonaffinity can be seen experimentally, although specific rheological models are required to quantify this connection.  Prior studies on dense suspensions \cite{tighe_model_2010,andreotti_shear_2012,lerner_unified_2012,woldhuis_fluctuations_2015,degiuli_unified_2015,ikeda_relaxation_2020,ikeda_nonaffine_2021}, foams and emulsions \cite{katgert_jamming_2013,boschan_stress_2017}, and immersed networks \cite{tighe_dynamic_2012,yucht_dynamical_2013,during_length_2014} have shown that a steady-state balance between externally applied power and the rate of dissipation by nonaffine rearrangement reveals phenomenological scaling relationships between the nonaffinity and  loss modulus. This has even been used to identify critical exponents, e.g., for networks near isostaticity \cite{yucht_dynamical_2013}.
Yet, many systems, including biopolymer networks such as the cellular cytoskeleton and extracellular matrix, are subjected to large and often transient applied stresses and strains; in cells and tissues, this gives rise to highly strain-dependent and typically power-law rheology \cite{trepat_universal_2007,mulla_origin_2019}, the origins of which are not yet fully understood. Given the potential for large energy-dissipating nonaffine rearrangement near the onset of tension-dominated rigidity \cite{onck_alternative_2005,huisman_monte_2008,sheinman_nonlinear_2012,sharma_strain-controlled_2016,shivers_nonlinear_2020}, one can assume that such rearrangements can lead to significant effects on network rheology in this regime. However, these effects remain poorly understood, especially in biopolymer or fiber systems with bending interactions, for which experimental measurement of nonaffinity has remained elusive.

Building on prior insights into the interplay between nonaffinity and energy dissipation, we identify a general relationship between the nonaffinity and measurable rheology of fluid-immersed networks. We find that the growth of nonaffine rearrangements near the strain stiffening transition drives a dramatic slowing down of stress relaxation in this regime. To explore the ensuing rheological consequences, we perform two- and three-dimensional simulations of prestrained disordered networks. We find that the longest relaxation time and nonaffinity both diverge as power laws with respect to distance to the stiffening transition. This leads to a set of scaling relations describing the relaxation modulus and nonaffinity near the critical strain, which we validate in simulations. We identify several experimentally testable predictions of this nonaffinity-dissipation relationship for a broad class of biopolymer and fiber systems.

\begin{figure}[htb!]
\centering
\includegraphics[width=1\columnwidth]{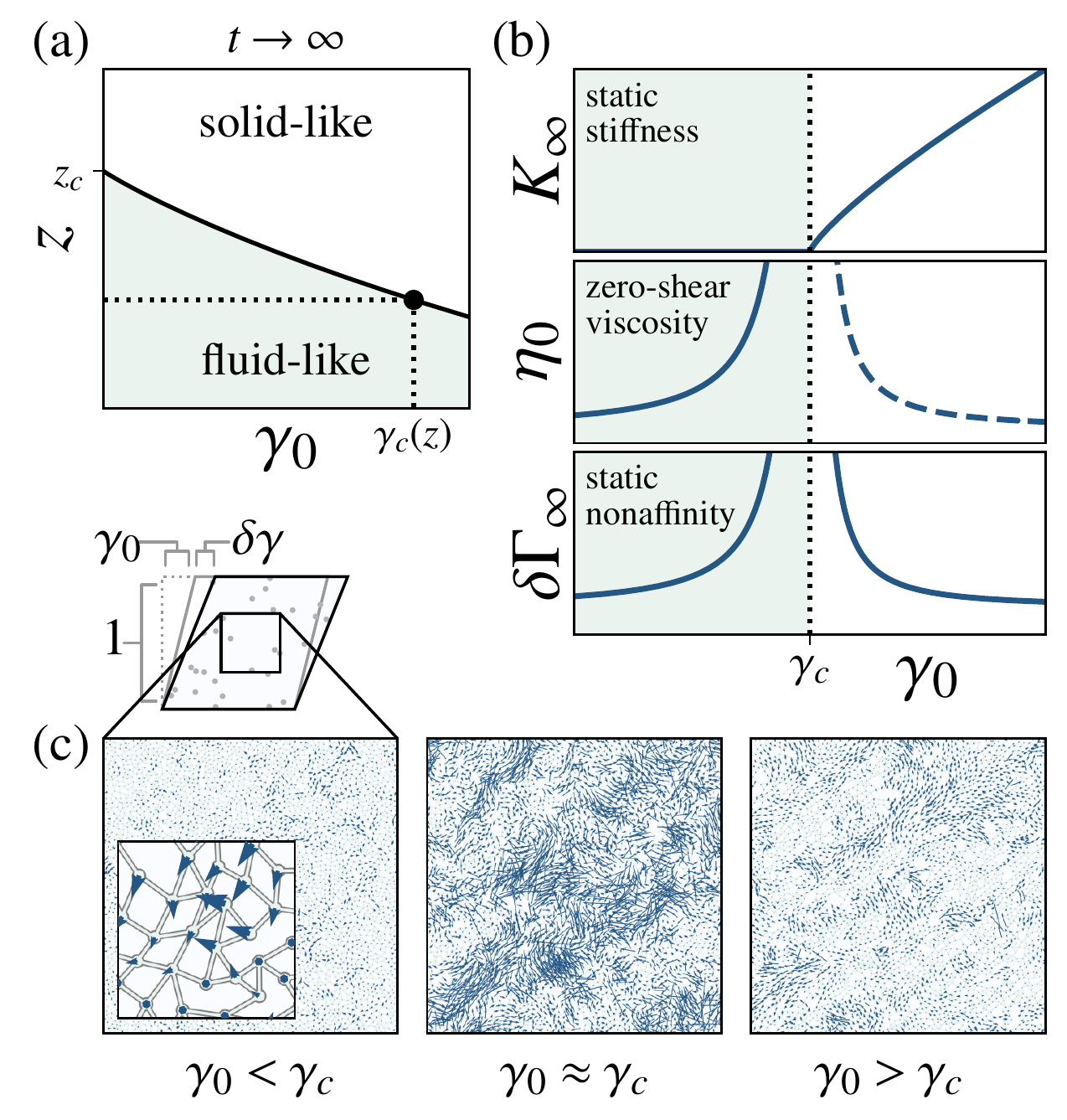}
\caption{ \label{fig1} (a) Immersed central-force spring networks with connectivity $z<z_c$ rigidify under shear strain $\gamma_0$ exceeding a $z$-dependent critical strain $\gamma_c$. (b) Rheological and kinematic features scale with the distance to the critical point, $|\gamma_0 - \gamma_c|$. At $\gamma_c$, the stiffness becomes nonzero, while the zero-shear viscosity and differential nonaffinity diverge.
(c) Energy stored by an \textit{affine} step strain $\delta\gamma$ is dissipated by microscopic nonaffine displacements $\mathbf u^{\mathrm{NA}}_i$, indicated here by arrows with uniformly scaled lengths.}
\end{figure}

We consider the overdamped dynamics of a $d$-dimensional system of $N$ particles with positions $\mathbf{r}_i$ interacting via a potential energy $U=f(\mathbf{r}_1,...,\mathbf{r}_N)$ \cite{particles_footnote}. These are immersed in a Newtonian fluid with velocity field $\mathbf{v}_f$, which imparts a drag force $\mathbf{f}_{d,i} = -\zeta(\dot{\mathbf{r}_i}-\mathbf{v}_f(\mathbf{r}_i))$ that balances the interaction force $\mathbf{f}_{p,i} = -\partial U/\partial \mathbf{r}_i$, such that $\mathbf{f}_{d,i}+\mathbf{f}_{p,i}=\mathbf{0}$. This ``free draining" description ignores long-range hydrodynamic interactions \cite{shankar_theory_2002}, which in our materials of interest can likely be neglected due to hydrodynamic screening.   We apply macroscopic shear strain $\gamma(t)$ via Lees-Edwards periodic boundary conditions \cite{lees_computer_1972} and assume that the fluid deforms affinely, such that $\mathbf{v}_f(\mathbf{r}_i)=r_{i,z} \dot{\gamma}(t) \mathbf{\hat{x}} $; this is the widely used ``affine solvent model'' \cite{durian_foam_1995,hatano_growing_2009,lerner_unified_2012,andreotti_shear_2012,yucht_dynamical_2013}.  For a given strain rate $\dot{\gamma}$, the macroscopic shear stress is $\sigma = \eta_f \dot{\gamma} + (2V)^{-1}\sum_{ij} f_{ij,x} r_{ij,z}$, in which $\eta_f$ is the fluid viscosity, $V$ is the system's volume, the sum is taken over all pairs of interacting particles $i$ and $j$, $\mathbf{f}_{ij}$ is the force on particle $i$ due to particle $j$, $\mathbf{r}_{ij} = \mathbf{r}_j - \mathbf{r}_i$, and $x$ and $z$ denote the flow and gradient directions, respectively.

Nonaffinity quantifies the reorganization required for a system initially in mechanical equilibrium (satisfying force balance or, equivalently, minimizing $U(\mathbf{r}_i)$) to re-equilibrate after a small affine perturbation \cite{tanguy_continuum_2002,didonna_nonaffine_2005}. Consider an energy-minimized system at some prestrain $\gamma_0$, to which we apply an instantaneous affine strain step $\delta\gamma$ yielding transformed particle positions $\mathbf{r}_{i,0}$ with, in general, a net force on each particle. Evolving the equations of motion until the forces are once again balanced, we find that the particles take on new positions $\mathbf{r}_{i,\infty}$ defining static nonaffine displacements $\mathbf{u}_{i,\infty}^{\mathrm{NA}} = \mathbf{r}_{i,\infty} - \mathbf{r}_{i,0}$, as sketched in Fig. \ref{fig1}c. These collectively define the static differential nonaffinity, $\delta\Gamma_\infty = (N \ell_0^2 \delta\gamma^2)^{-1}\sum_i\| \mathbf{u}_{i,\infty}^{\mathrm{NA}}\|^2$. As noted earlier, in response to even small perturbations, amorphous materials near marginal stability tend to undergo large-scale rearrangement signaled by large $\delta\Gamma_{\infty}$.

We consider discrete elastic networks of central-force springs with stretching rigidity $\mu$ and angular springs with bending rigidity $\kappa$, prepared as described in Supplemental Material \cite{supplemental_material}. We focus on subisostatic networks, i.e.\ those with average connectivity $z$ (number of bonds connected to each node) below Maxwell's $d$-dependent isostatic point $z_{c} = 2d$  \cite{maxwell_calculation_1864}. For biopolymer networks, $z$ is generally between 3 and 4 \cite{lindstrom_biopolymer_2010}, far below $z_c$ in $d=3$. The linear elastic moduli of subisostatic networks, in the static ($t\to\infty$) limit, are proportional to $\kappa$. For $\kappa = 0$, they are thus \textit{floppy} in the small strain limit \cite{alexander_amorphous_1998} but can transition to a tension-stabilized rigid regime under finite applied strain. We select simple shear prestrain $\gamma_0$ as the rigidity control variable; in this case, static ($t\to\infty$) solid-like behavior develops when $\gamma_0$ reaches the $z$-dependent critical strain $\gamma_c$ \cite{wyart_elasticity_2008}, as shown in Fig. \ref{fig1}a. For $N, V \to \infty$, as $\gamma_0$ approaches $\gamma_c$, the system's zero-shear viscosity and nonaffinity diverge, as sketched in Fig. \ref{fig1}b.

\begin{figure}[htb!]
\centering
\includegraphics[width=1\columnwidth]{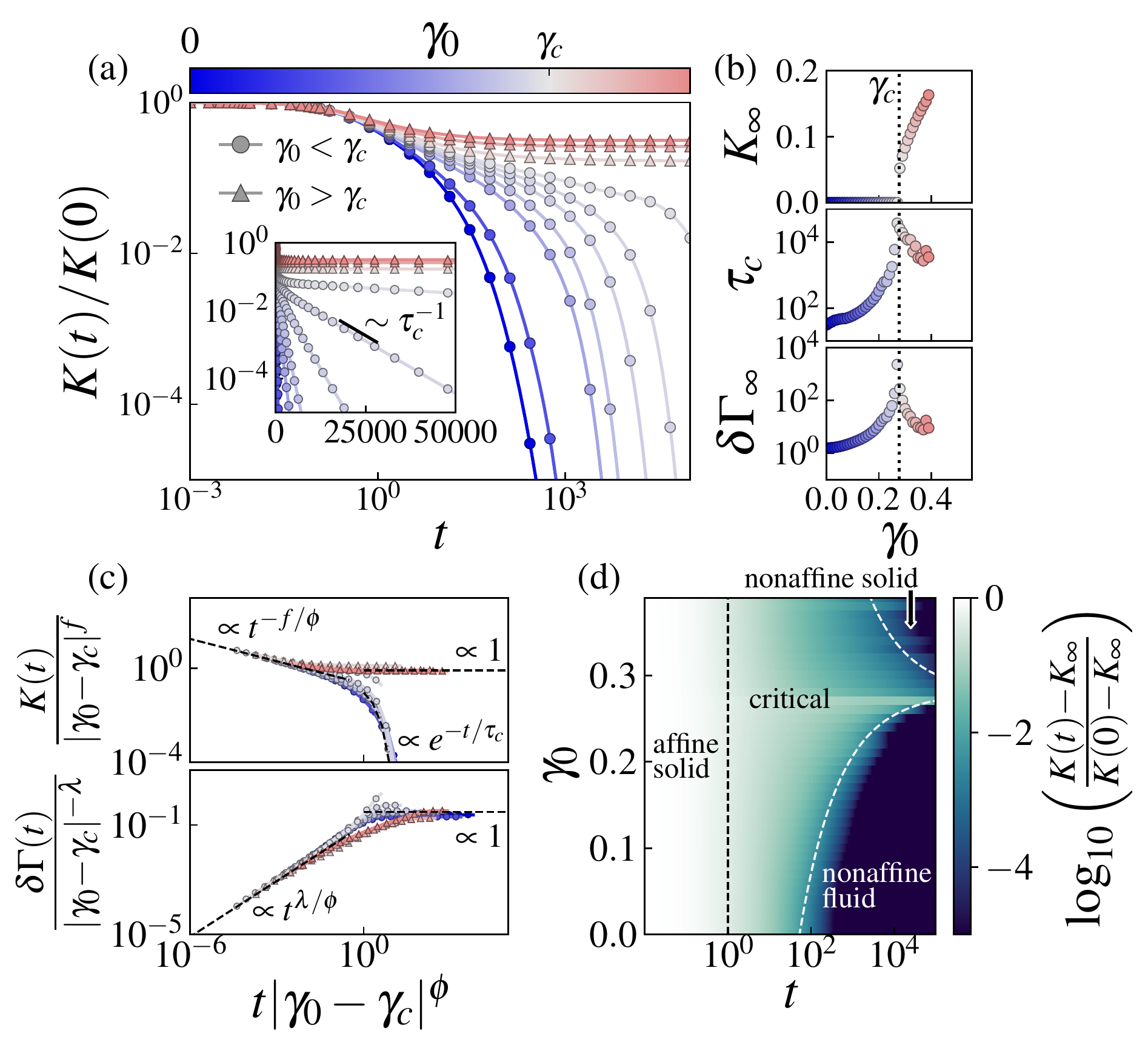}
\caption{\label{fig2} 
When an immersed spring network at prestrain $\gamma_0$ is subjected to an instantaneous, infinitesimal strain increment $\delta\gamma$, (a) the relaxation modulus $K(t)$ decays to the static ($t\to\infty$) value $K_\infty$ with a slowest relaxation time $\tau_c$ that (b) diverges at a critical prestrain  $\gamma_0=\gamma_c$ along with the static nonaffinity $\delta\Gamma_\infty$. (c) $K(t)$ and $\delta\Gamma(t)$ collapse according to the Widom-like scaling forms of Eqs. \ref{widomstiffness} and \ref{widomnonaff}, with  exponents $f = 0.7$, $\phi = 2.2$, and $\lambda = 1.5$. (d) Viscoelastic regimes on a $(\gamma_0,t)$ phase diagram. Dashed white curves show $t\propto|\gamma_0-\gamma_c|^{-\phi}$. Here, $N=6400$, $z=3.5$, $\tilde{\kappa}=0$ and $\delta\gamma = 10^{-4}$.
}
\end{figure}

In Fig. \ref{fig1}c, we plot static nonaffine displacement vectors for a representative network with $z=3.5$ under varying prestrain. The nonaffine displacements are largest at the critical strain $\gamma_c$ corresponding to the stiffening transition \cite{onck_alternative_2005} (see Fig. \ref{fig2}b). Although the corresponding maximum in the static nonaffinity $\delta\Gamma_\infty$ provides a clear signal of the critical point in simulations, its measurement in experiments, often by tracking embedded tracer particles \cite{wen_local_2007,liu_visualizing_2007,basu_nonaffine_2011}, is challenging and limited in precision. Ideally, one could measure nonaffinity by relating it to more experimentally accessible quantities, such as the viscoelastic moduli. As noted earlier, such a relationship exists due to energy conservation: at steady state, the power injected into the system by the externally applied stress balances the power dissipated by the nonaffine rearrangement \cite{tighe_model_2010,andreotti_shear_2012,lerner_unified_2012,woldhuis_fluctuations_2015,degiuli_unified_2015,ikeda_relaxation_2020,ikeda_nonaffine_2021,katgert_jamming_2013,boschan_stress_2017,tighe_dynamic_2012,yucht_dynamical_2013,during_length_2014}.

A similar power balance relates nonaffinity and viscoelasticity beyond the linear regime. Consider an energy-minimized configuration under prestrain $\gamma(t\le 0)=\gamma_0$ subjected to a superimposed oscillatory strain of amplitude $\delta\gamma$ and frequency $\omega$, such that  $\gamma(t) = \gamma_0 + \delta\gamma\sin(\omega t)$ for $t > 0$. After an initial transient regime, the stress steadily oscillates as $\sigma(t) = \sigma_0 + \delta\sigma \sin(\omega t + \theta)$, with amplitude $\delta\sigma$ and phase shift $\theta$. Equivalently, $\sigma(t) = \sigma_0 + \delta\gamma\left(K'\sin(\omega t) + K''\cos(\omega t)\right)$, in which $K'(\gamma_0,\omega)=(\delta\sigma/\delta\gamma)\cos\theta$ and $K''(\gamma_0,\omega)=(\delta\sigma/\delta\gamma)\sin\theta$ are the frequency-dependent differential storage and loss moduli. For small $\delta\gamma$ \cite{dagois-bohy_softening_2017}, particles adopt elliptical trajectories $\mathbf{p}(t) = \mathbf{r}_i(t) - \mathbf{r}_{i,0}$ combining affine and nonaffine components $\mathbf{p}_i^{\mathrm{A}}(t) = \mathbf{u}_i^{\mathrm{A}}(\omega) \sin (\omega t + \theta^\mathrm{A})$ and $\mathbf{p}_i^{\mathrm{NA}}(t) = \mathbf{u}_i^{\mathrm{NA}}(\omega) \sin (\omega t + \theta^\mathrm{NA})$, with $\mathbf{p}(t) = \mathbf{p}^{\mathrm{A}}(t) + \mathbf{p}^{\mathrm{NA}}(t)$. The nonaffine displacement vectors collectively define the frequency-dependent nonaffinity, $\delta\Gamma(\omega)= (N\ell_0^2\delta\gamma^2)^{-1}\sum_i \| \mathbf{u}_i^\mathrm{NA}(\omega) \|^2 $, in which $\ell_0$ is a characteristic length scale, e.g.\ the typical spring length. The drag force on each particle is proportional to its velocity relative to the fluid, $\partial \mathbf{p}_i^{\mathrm{NA}}/\partial t=\omega\mathbf{u}_i^{\mathrm{NA}}(\omega)\cos(\omega t +\theta^{\mathrm{NA}})$. Averaged over each cycle, the external power input $P_\mathrm{in}=\frac{1}{2}V\omega d\gamma^2\left(K''-\eta_f \omega\right)$ balances the total power output by nonaffine work, $P_\mathrm{out}=\frac{1}{2} N \omega^2 \zeta \ell_0^2 \delta\gamma^2 \delta\Gamma(\omega)$ \cite{supplemental_material}. Thus, for any prestrain, we can express the differential dynamic viscosity $\eta'(\omega) = K''(\omega)/\omega$ in terms of the frequency-dependent nonaffinity as
\begin{equation} \label{freqdependent}
\eta'(\omega)-\eta_f = \rho \zeta \ell_0^2 \delta\Gamma(\omega)
\end{equation}
in which $\rho = N/V$ is the particle number density. For $\omega\to0$, this relates the zero-shear differential viscosity $\eta_0 = \lim_{\omega\to 0} \eta'(\omega)$ and the static nonaffinity $\delta\Gamma_\infty = \lim_{\omega\to 0} \delta\Gamma(\omega)$ as
\begin{equation} \label{zeroshear}
\eta_0 - \eta_f = \rho \zeta \ell_0^2 \delta\Gamma_\infty.
\end{equation}
The latter indicates that, for a free-draining suspension with fluid viscosity $\eta_f$, the increase in zero-shear viscosity due to the presence of interacting particles, $\eta_0-\eta_f$, is proportional to the fluid-independent static nonaffinity associated with the particle arrangement, $\delta\Gamma_\infty$. As this relationship is independent of $U$, it applies to a wide range of systems including, as we will later demonstrate, networks of bending-resistant filaments and soft sphere suspensions near jamming.

We now consider the effects of these relationships on the dynamic response of a strained network to an instantaneous strain step. To a relaxed system at prestrain $\gamma_0$, we apply an affine strain step $\delta\gamma$, such that $\gamma(t) = \gamma_0 + \delta\gamma$ for $t \ge 0$. The particles adopt nonaffine trajectories $\mathbf{u}_i^{\mathrm{NA}}(t) = \mathbf{r}_i(t) - \mathbf{r}_i(0)$ that collectively define the relaxation nonaffinity $\delta\Gamma(t) = (N\ell_0^2 \delta\gamma^2)^{-1}\sum_i {\|\mathbf{u}_i^\mathrm{NA}(t) \|^2}$,
for which $\delta\Gamma_\infty = \lim_{t \to \infty} \delta\Gamma (t)$. We measure the corresponding change in shear stress $\delta\sigma(t)=\sigma(t)-\sigma_0$ and compute the  differential relaxation modulus $K(t) = \delta\sigma/\delta\gamma$ and differential zero-shear viscosity $\eta_0 - \eta_f = \int_0^\infty (K(t) - K_\infty) dt$, in which the static differential modulus is $K_{\infty}=\lim_{t \to \infty} K(t)$. Note that, for $\gamma_0 = 0$,  $K$ and $\delta\Gamma$ are the linear relaxation modulus $G(t) = \lim_{\gamma_0 \to 0} K(t)$ and linear nonaffinity $\Gamma(t) = \lim_{\gamma_0 \to 0} \delta\Gamma(t)$.

Because the static nonaffinity $\delta\Gamma_\infty$ diverges at the critical strain, Eq. \ref{zeroshear} implies that we should observe an equivalently diverging zero-shear viscosity and associated diverging slowest relaxation time.  In Fig. \ref{fig2}a, we plot stress relaxation curves for a single two-dimensional network with $z=3.5$, with infinitesimal step strains applied over a range of prestrains $\gamma_0$ containing $\gamma_c$. The normalized relaxation modulus $K(t)/K(0) = \delta\sigma(t)/\delta\sigma(0)$ decays to its equilibrium value $K_\infty /K(0)$ with a $\gamma_0$-dependent slowest relaxation time $\tau_c$ (calculated as described in Supplemental Material \cite{supplemental_material}), which is plotted in Fig. \ref{fig2}b as a function of $\gamma_0$ along with the corresponding static nonaffinity $\delta\Gamma_\infty$, and static differential modulus $K_\infty = \delta\sigma_\infty/\delta\gamma$. Maxima in $\tau_c$ and $\delta\Gamma_\infty$ occur at the critical strain, where $K_\infty$ becomes nonzero. We assign the exponent $\phi$ to the scaling of $\tau_c$ with $|\gamma_0-\gamma_c|$ and, following Ref. \cite{sharma_strain-controlled_2016}, assign $\lambda$ to $\delta\Gamma_\infty$ and $f$ to  $K_\infty$.

The relaxation modulus exhibits power-law decay over a range of times extending from the microscopic relaxation time $\tau_0=\zeta\ell_0/\mu=1$ to a characteristic slow timescale governed by the distance from the critical strain, $\tau_c = |\gamma_0-\gamma_c|^{-\phi}$. Within this regime, the relaxation modulus is a function of the ratio $t/\tau_c$. Beyond $\tau_c$, we expect the static critical behavior, i.e.\ $K_\infty \propto |\gamma_0-\gamma_c|^f$ for $\gamma_0 \ge \gamma_c$. Thus $K(t)$ should obey the scaling form
\begin{equation} \label{widomstiffness}
\begin{split}
K(t) =|\gamma_0-\gamma_c|^f \mathcal{F}_{\pm}\left(t|\gamma_0-\gamma_c|^{\phi}\right) 
\end{split}
\end{equation}
in which the branches of the scaling function $\mathcal{F}_\pm(x)$ correspond to regimes above and below the critical strain. When $x \gg 1$, $\mathcal{F}_+(x)\sim \mathrm{constant}$ and $\mathcal{F}_-(x) \sim \exp(-x)$, implying $K(t) \sim |\gamma_0-\gamma_c|^f$ above $\gamma_c$ and $K(t)\sim |\gamma_0-\gamma_c|^f \exp(-t|\gamma_0-\gamma_c|^{\phi})$ below $\gamma_c$. When $x \ll 1$, $K(t)$ remains finite and thus must be independent of $|\gamma_0-\gamma_c|$, so $\mathcal{F}_\pm(x)\sim x^{-f/\phi}$. Therefore, for $\gamma_0 = \gamma_c$, the relaxation modulus is predicted to decay as $K(t) \propto t^{-f/\phi}$.

Near $\gamma_c$, the differential nonaffinity is controlled by the same diverging timescale $\tau_c$, yet should eventually display the static critical behavior $\delta\Gamma \propto |\gamma_0-\gamma_c|^{-\lambda}$. We thus expect
\begin{equation} \label{widomnonaff}
\begin{split}
\delta\Gamma(t) =|\gamma_0-\gamma_c|^{-\lambda}\mathcal{G}_\pm\left(t|\gamma_0-\gamma_c|^\phi\right)
\end{split}
\end{equation}
in which, for $x \gg 1$, $\mathcal{G}_+(x)\sim \mathrm{constant}$ and $\mathcal{G}_-(x) \sim \mathrm{constant}$. Because $\delta\Gamma(t)$ remains finite when $x \ll 1$, $\mathcal{G}_\pm(x)\sim x^{\lambda/\phi}$. Thus for  $\gamma_0 = \gamma_c$, the nonaffinity grows as $\delta\Gamma(t) \propto t^{\lambda/\phi}$. We observe excellent collapse of $K(t)$ and $\delta\Gamma(t)$ according to these scaling forms with exponents $f = 0.7$, $\phi = 2.2$, and $\lambda = 1.5$, as shown in Fig. \ref{fig2} \cite{exponent_footnote}.

\begin{figure}[htb!]

\centering
\includegraphics[width=1\columnwidth]{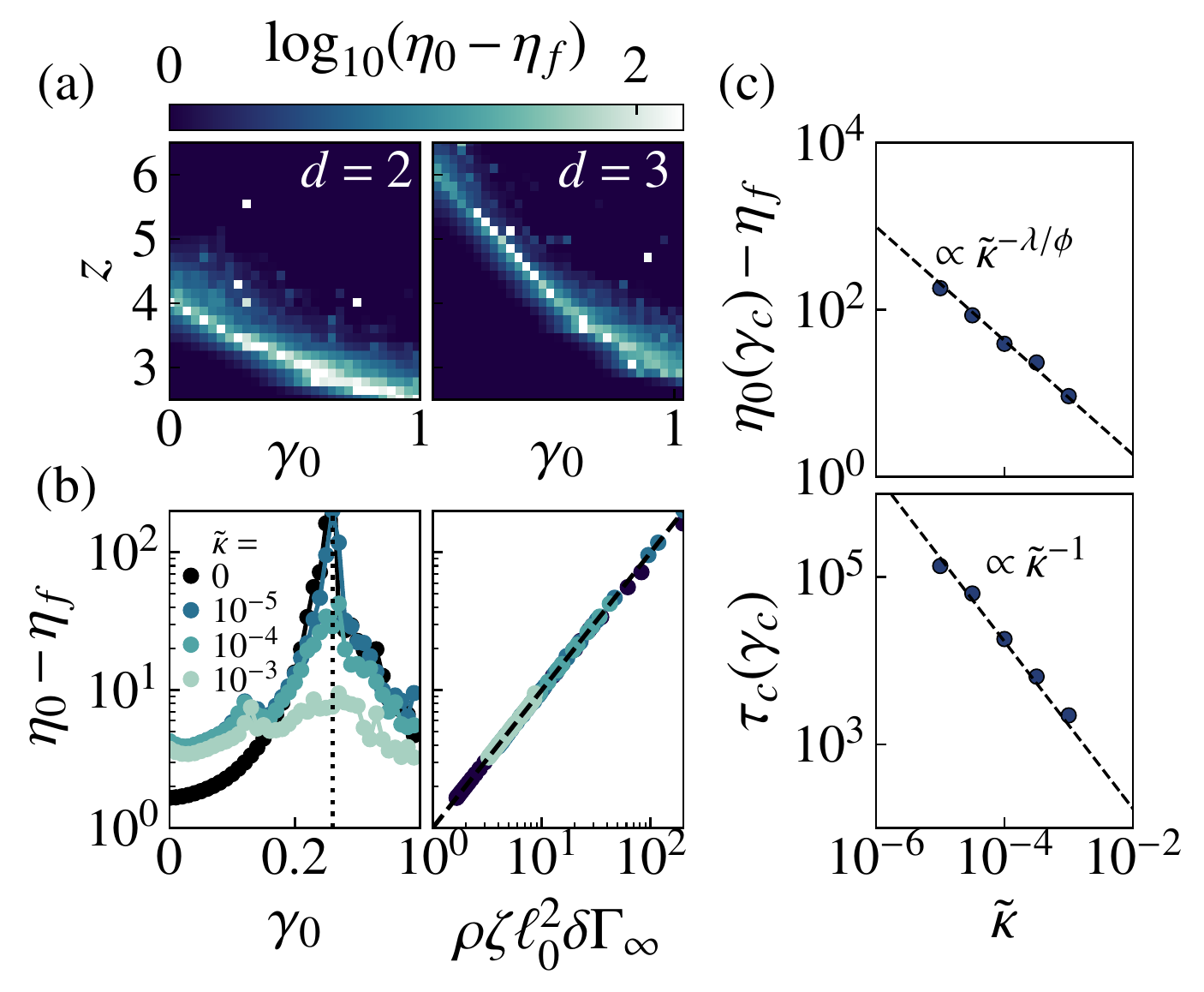}
\caption{\label{fig3} In immersed networks, the zero-shear viscosity $\eta_0-\eta_f$ is maximized at the $z$-dependent critical strain $\gamma_c$, mirroring the nonaffinity. Here, $N(d=2)=10000$, $N(d=3)=8000$, and $\tilde{\kappa} = 0$. (b)  Finite bending rigidity $\tilde{\kappa}$ suppresses the divergence of $\eta_0 -\eta_f$ at $\gamma_c$, yet Eq. \ref{zeroshear} remains satisfied. Here, $N = 1600$. (c) Peaks in the viscosity and slowest relaxation time decay with $\tilde{\kappa}$ as predicted by Eqs. \ref{eta_kappa} and \ref{tau_kappa}.} 
\end{figure}

We next test Eq. \ref{zeroshear}, which relates the independently measured static nonaffinity $\delta\Gamma_{\infty}$ and zero-shear viscosity $\eta_0$. In Fig. \ref{fig3}a, we demonstrate that, like $\delta\Gamma_{\infty}$, $\eta_0$ is maximized at the finite-strain phase boundary between the statically floppy and rigid regimes. In Fig. \ref{fig3}b, we plot $\eta_0-\eta_f$ for networks with varying dimensionless bending rigidity $\tilde{\kappa}$ and observe, in agreement with Eq. \ref{zeroshear}, a divergence in $\eta_0-\eta_f$ at the critical strain that is suppressed by increasing $\tilde{\kappa}$, which acts as a stabilizing field \cite{sharma_strain-controlled_2016}. In Supplemental Material \cite{supplemental_material}, we verify that the same nonaffinity-viscosity relationship applies in dense suspensions of frictionless soft spheres, in which $\eta_0$ diverges at a critical volume fraction $\phi_c$.

The aforementioned power balance connects the static scaling exponents, $f$ and $\lambda$, and the dynamic exponent, $\phi$. At $\gamma_c$, the relaxation modulus decays as $K(t)\sim t^{-f/\phi}$ and the nonaffinity grows as $\delta\Gamma(t)\sim t^{\lambda/\phi}$, so the corresponding frequency dependence of the complex modulus and nonaffinity must be $K^*(\omega)\sim\omega^{f/\phi}$ and $\delta\Gamma(\omega)\sim\omega^{-\lambda/\phi}$. The former implies $\eta'(\omega)\sim \omega^{f/\phi-1}$, hence Eq. \ref{freqdependent} implies $\phi = f + \lambda$.
Consequently, the static scaling of the stiffness and nonaffinity controls $\phi$ and, by extension, the exponents $f/\phi$ and $\lambda/\phi$ describing the system's stress relaxation and time-dependent rearrangement, as prior work has noted for networks near isostaticity \cite{yucht_dynamical_2013}. Alternatively, we can rationalize this finding with a more qualitative argument: the relaxation time of large structural rearrangements scales with the ``size'' of these rearrangements, i.e.\ $\delta\Gamma_{\infty}\sim |\Delta\gamma|^{-\lambda}$, divided by the magnitude of their driving force, proportional to $K_{\infty}\sim|\Delta\gamma|^f$, hence $\tau_c\sim|\Delta\gamma|^{-(\lambda+f)}$, with units set by the fluid viscosity. This relationship implies that the dynamic exponent $\phi$ is identical to the exponent describing the critical coupling to the bending rigidity, defined in Ref. \cite{sharma_strain-controlled_2016}. Therefore, at $\gamma_c$, the excess zero-shear viscosity should scale with $\tilde{\kappa}$ just as $\delta\Gamma_\infty$ does \cite{shivers_scaling_2019},
\begin{equation} \label{eta_kappa}
\eta_0(\gamma_c) - \eta_f \propto \tilde{\kappa}^{-\lambda/\phi},
\end{equation} and the slowest relaxation time should scale as 
\begin{equation} \label{tau_kappa}
\tau_c(\gamma_c) \propto \tilde{\kappa}^{-1},
\end{equation}
independently of the critical exponents. These relationships are satisfied in simulations, as shown in Fig. \ref{fig3}c.

Several of our predictions are experimentally testable. For example, the exponents $f$ and $\phi$ (and thus $\lambda$) can be estimated via quasistatic strain-controlled rheology, as shown using reconstituted collagen networks in Ref. \cite{sharma_strain-controlled_2016}, after which the predicted scaling of $\eta'(\omega)$ for networks at $\gamma_c$, $\eta'(\omega)\propto \omega^{f/\phi-1}$, can be tested via small-amplitude oscillatory rheology at finite prestrain, i.e. $\gamma(t) = \gamma_0 + \delta\gamma\sin(\omega t)$. In addition, $\eta_0$ and $\tau_c$ can be determined via step-strain stress relaxation tests, with $\gamma(t) = \gamma_0$ for $t < 0$ and $\gamma(t) = \gamma_0 + \delta\gamma$ for $t \ge 0$, allowing for tests of the predictions $\eta_0 - \eta_f \propto |\gamma_0 - \gamma_c|^{f-\phi}$ and $\tau_c \propto |\gamma_0 - \gamma_c|^{-\phi}$. In the same manner, the predicted dependence of $\eta_0$ and $\tau_c$ on $\tilde{\kappa}$ can be tested using reconstituted collagen networks of varying concentration $c$, for which prior work has shown $\tilde{\kappa} \propto c$ \cite{sharma_strain-controlled_2016,jansen_role_2018}; hence, one would expect $\eta_0(\gamma_c)-\eta_f \propto c^{-\lambda/\phi}$ and $\tau_c \propto c^{-1}$. 

In conclusion, we have demonstrated that a fundamental quantitative relationship between nonaffine fluctuations and excess viscosity controls the rheology of immersed networks near the onset of rigidity. Consequently, the phase boundary for strain-induced stiffening in subisostatic networks is accompanied by a diverging excess viscosity. Applying prestrain to such networks thus produces a dramatic slowing of stress relaxation that is nonetheless quantitatively predictable from quasistatic nonaffine fluctuations. We provided experimentally testable predictions for the dynamics of networks near $\gamma_c$, with broad implications for the rheology of biological materials. To emphasize the generality of the nonaffinity-viscosity relationship, we showed that it fully captures the diverging zero-shear viscosity in suspensions of soft frictionless spheres near jamming in two and three dimensions \cite{supplemental_material}.

There is widespread interest in the rational design of materials with tunable viscoelasticity \cite{raffaelli_stress_2021,lin_tunable_2016,chaudhuri_hydrogels_2016}. This generally involves adjusting aspects of a material's preparation, such as polymer concentration or particle volume fraction. However, the connection between nonaffine fluctuations and excess viscosity implies that, in fiber networks, one can generate dramatic changes in stress relaxation dynamics by simply applying external strain, without changing the underlying network structure. This suggests other avenues for tuning the dynamics of stress relaxation; for example, embedded force-generating components can drive macroscopic stiffening transitions and thus precisely control the nonaffinity \cite{sheinman_actively_2012}. Examples include cytoskeletal molecular motors \cite{koenderink_active_2009,broedersz_molecular_2011,wang_active_2012}, contractile cells \cite{jansen_cells_2013,ronceray_fiber_2016,han_cell_2018}, and inclusions driven to shrink by varying temperature \cite{chaudhary_thermoresponsive_2019} or rearrange under applied magnetic fields \cite{chaudhary_exploiting_2020}.

Additional work will be needed to characterize the effects of finite system size on nonaffinity-induced critical slowing down near the onset of rigidity. For networks at the critical strain with correlation length exponent $\nu$, we expect $\tau_c(\gamma_c)\propto L^{\phi/\nu}$ and $\eta_0(\gamma_c)-\eta_f \propto L^{\lambda/\nu}$ \cite{supplemental_material}, suggesting additional ways to identify $\nu$ and test the previously proposed hyperscaling relation, $\nu = (f+2)/d$ \cite{shivers_scaling_2019}. Other areas to investigate include the effects of hydrodynamic interactions near the critical strain, as these both increase nonaffinity near isostaticity \cite{dennison_viscoelastic_2016} and couple with nonaffinity to produce an additional intermediate-frequency viscoelastic regime at small strains \cite{head_nonaffinity_2019}, and the effects of finite temperature: the Green-Kubo relations tie the stationary stress correlations to the zero-shear viscosity \cite{levesque_computer_1973,visscher_dynamic_1994} and thus to the athermal static nonaffinity.  We note also that the association of diverging nonaffine fluctuations with the onset of rigidity, coupled with their microscopic role in slowing stress relaxation, may account for prior observations of slow dynamics in disordered materials such as fractal colloidal gels \cite{larsen_microrheology_2008,aime_power_2018} and crowded, prestressed living cells \cite{fabry_scaling_2001,bursac_cytoskeletal_2005,trepat_universality_2008,pritchard_mechanics_2014}.
 Finally, it remains to be seen whether connections between nonaffinity and slowing down might provide insight into the glass transition \cite{stevenson_shapes_2006,leonforte_inhomogeneous_2006,lubchenko_theory_2007,brambilla_probing_2009,ballauff_residual_2013,ikeda_unified_2012,bonn_yield_2017}.

\begin{acknowledgments}
This work was supported in part by the National Science Foundation Division of Materials Research (Grant No.\ DMR-2224030) and the National Science Foundation Center for Theoretical Biological Physics (Grant No.\ PHY-2019745). J.L.S. acknowledges additional support from the Lodieska Stockbridge Vaughn Fellowship.
\end{acknowledgments}

\bibliographystyle{apsrev}
\bibliography{Bibliography, supp}{}

\begin{thebibliography}{104}
\bibitem[{exp()}]{exponent_footnote}
\bibinfo{note}{We note that the exponents $f$, $\phi$, and $\lambda$ here,
  which describe the scaling of various quantities with respect to $|\gamma_0 -
  \gamma_c|$ for $z < z_c$, need not be the same as the corresponding exponents
  with respect to $|z - z_c|$ for $\gamma_0 = 0$ in, e.g., Refs.
  \cite{broedersz_criticality_2011, yucht_dynamical_2013}. In general, it has
  been observed that the values of the critical exponents for the isostatic
  point and strain-controlled transition are different, i.e. $f_\gamma \neq
  f_z$ and $\phi_\gamma \neq \phi_z$.}
\end{thebibliography}

\end{document}


\title{Supplemental Material -- Strain-controlled critical slowing down in the rheology of disordered networks}


\author{Jordan L.\ Shivers}
\email{jshivers@uchicago.edu}
\altaffiliation[Present affiliation: ]{James Franck Institute and Department of Chemistry, University of Chicago, Chicago, Illinois 60637, USA}
\affiliation{Department of Chemical and Biomolecular Engineering, Rice University, Houston, TX 77005, USA}
\affiliation{Center for Theoretical Biological Physics, Rice University, Houston, TX 77005, USA}
\author{Abhinav Sharma}
\affiliation{Institute of Physics, University of Augsburg, 86159 Augsburg, Germany}
\affiliation{Leibniz-Institut für Polymerforschung Dresden, Institut Theorie der Polymere, 01069 Dresden, Germany}
\author{Fred C.\ MacKintosh}
\affiliation{Department of Chemical and Biomolecular Engineering, Rice University, Houston, TX 77005, USA}
\affiliation{Center for Theoretical Biological Physics, Rice University, Houston, TX 77005, USA}
\affiliation{Department of Chemistry, Rice University, Houston, TX 77005, USA} 
\affiliation{Department of Physics \& Astronomy, Rice University, Houston, TX 77005, USA}

\maketitle
\hypersetup{hidelinks}

\renewcommand{\baselinestretch}{0}\normalsize
\tableofcontents
\renewcommand{\baselinestretch}{1.0}\normalsize

\section{Disordered network model}

We consider the behavior of a bond-bending network \cite{arbabi_elastic_1988}  in a Newtonian solvent, in which drag forces act only on network nodes. We derive network structures with initial connectivity $z_0\approx 6$ in 2D and $z_0 \approx 10$ in 3D from the contact networks of dense packings of soft spheres, which are generated using protocols described in prior work \cite{wyart_elasticity_2008,tighe_dynamic_2012,shivers_scaling_2019}. We then reduce the connectivity $z$ to the desired value by selectively removing bonds randomly chosen from the set of nodes with the highest coordination number, yielding a network with a relatively homogeneous connectivity distribution \cite{wyart_elasticity_2008}. The energy of the network is
\[
U = \frac{\mu}{2} \sum_{ij} \frac{\left(\ell_{ij}-\ell_{ij,0}\right)^2}{\ell_{ij,0}} + \frac{\kappa}{2} \sum_{ijk} \frac{\left(\theta_{ijk}-\theta_{ijk,0}\right)^2}{\ell_{ijk,0}}
\]
in which $\mu$ is the bond stretching stiffness (units of energy $/$ length), $\kappa$ is the bending rigidity (units of energy $\times$ length) acting between adjacent bonds, the instantaneous and rest lengths of bond $ij$ are $\ell_{ij} = \mathbf{r}_j - \mathbf{r}_i$ and $\ell_{ij,0}$, the instantaneous and rest angle between bonds $ij$ and $jk$ are $\theta_{ijk}$ and $\theta_{ijk,0}$, and $\ell_{ijk,0} = (\ell_{ij,0} + \ell_{jk,0})/2$. We define the rest lengths and angles such that $U(\gamma_0=0)=0$.

The node dynamics follow the over-damped, zero-temperature Langevin equation,  
\[
-\frac{\partial U}{\partial \mathbf{r}_i} - \zeta \left(\frac{d \mathbf{r}_i}{dt}-\mathbf{v}_f(\mathbf{r}_i)\right) = \mathbf{0}
\]
in which $\zeta$ is the drag coefficient. Here, $\mathbf{v}_f(\mathbf{r}_i)$ denotes the velocity of the solvent at the position of node $i$. Note that we are using a free-draining \cite{visscher_dynamic_1994} approximation and thus ignoring hydrodynamic interactions between nodes. We integrate this equation using the Euler method with timestep $\Delta t = 10^{-3}$.  For convenience, we set $\mu = \zeta = 1$ and vary $\tilde{\kappa} = \kappa/(\mu \ell_0^2)$. Note that for $\tilde{\kappa} = 0$, the characteristic microscopic relaxation time is $\tau_0 = \zeta \ell_0 /\mu$.

\section{Stress relaxation at finite strain}

We first obtain the minimum energy configuration of the network at applied shear strain $\gamma = \gamma_0$ using the conjugate gradient method with a stopping criterion of $f_\mathrm{max}<10^{-12}$, in which $f_\mathrm{max}$ is the magnitude of the largest net force on any node. Then, we apply a small, instantaneous \textit{affine} shear strain step $\delta \gamma = 10^{-3}$, such that the strain becomes $\gamma = \gamma_0 + \delta\gamma$. Taking this as the initial state and assuming the solvent is immobile ($\mathbf{v}_f =\mathbf{0}$) in this case, we allow the system to evolve according to the equations of motion. We measure the shear stress $\sigma( t)$ as a function of time as the system evolves and compute the differential relaxation modulus,
\[
 K(\gamma_0,  t) = \lim_{\delta\gamma\to0}\frac{\sigma(\gamma_0 + \delta\gamma,  t) - \sigma(\gamma_0,  t\to\infty)}{\delta\gamma}
\]
Note that $ K_{\mathrm{aff}}(\gamma_0) \equiv K(\gamma_0, t=0)$ corresponds to the affine differential modulus, and the system eventually settles to the equilibrium (long-time) differential modulus $ K_{\infty}(\gamma_0) \equiv  K(\gamma_0,  t\to \infty)$, equivalent to that measured under quasistatic shear.

\section{Small-amplitude oscillatory shear at finite strain}

For small-amplitude oscillatory shear near the critical strain, the system exhibits power-law scaling of the dynamic moduli over a range of frequencies bounded on the lower end by the critical characteristic frequency $\omega_c=|\gamma_0-\gamma_c|^\phi$, governed by the proximity to the critical strain, and on the upper end by the characteristic frequency $\omega_0 \approx 1$ above which the network behaves as a solid. We assume that the ratio $\omega/\omega_c = \omega|\gamma_0-\gamma_c|^{-\phi}$ governs the mechanics for a particular strain, in which case the differential modulus takes on the scaling form
\[
\begin{split}
 K'(\omega) =|\gamma_0-\gamma_c|^f \mathcal{H}_\pm\left(\omega|\gamma_0-\gamma_c|^{-\phi}\right) \\
\end{split}
\]
in which, for $x\gg 1$, $\mathcal{H}_-(c) \sim x^2 $ and $\mathcal{H}_+(x) \sim \mathrm{constant}$, while for $x \gg 1$ we must have $\mathcal{H}_{\pm}(x)\propto x^{f/\phi}$ since $ K'(\omega)$ remains finite.

Since at the critical strain we have $\delta\Gamma(\omega)\propto \omega^{-\lambda/\phi}$ and $ K^*(\omega)\propto f/\phi$ (i.e.\ $\eta^*(\omega)\propto^{f/\phi-1}$), the relation above implies $f/\phi - 1 = -\lambda/\phi$, or
\[
f = \phi - \lambda
\]
as we find for the stress relaxation case.
Note that in Ref. \cite{yucht_dynamical_2013}, Yucht et al. made essentially the same argument relating the scaling behavior of the linear loss modulus and nonaffinity for networks near the isostatic point.

\section{Power balance}

In the steadily oscillating regime (long after initiating the small-amplitude oscillatory shear), the power injected in the external application of strain, averaged over a single cycle, is
\[
\begin{split}
P_\mathrm{in} & =  V \frac{\omega}{2 \pi}\int_{t_0}^{t_0 + 2\pi/\omega}\dot{\gamma} \left(\delta\sigma-\eta_f\dot{\gamma}\right)  dt \\ & = V \frac{\omega}{2 \pi} \int_{t_0}^{t_0 + 2\pi/\omega} \omega \delta\gamma \cos\omega t\left(\delta\gamma\left[K'\sin\omega t + K''\cos\omega t\right] - \eta_f \omega \delta\gamma \cos \omega t \right)dt \\ & = \frac{1}{2} V \omega \delta\gamma^2 (K'' - \eta_f\omega )
\end{split}
\]
The nonaffine displacement of node $i$ is $\mathbf{p}_i^{\mathrm{NA}}(t) = \mathbf{u}_i^\mathrm{NA}(\omega) \sin \left(\omega t + \theta^{\mathrm{NA}}\right)$, and the nonaffine velocity is $\partial\mathbf{p}_i^{\mathrm{NA}}/\partial t = \omega \mathbf{u}_i^\mathrm{NA}(\omega)  \cos \left(\omega t + \theta^{\mathrm{NA}}\right)$. The system-wide instantaneous nonaffinity is $\delta\Gamma_i(\omega) = \sum_i \left\|\mathbf{u}_i^\mathrm{NA}(\omega)\right\|^2\sin^2(\omega t + \theta^\mathrm{NA})/(\ell_0^2 \delta\gamma^2)$.
The power output, averaged over a single cycle, in dragging the nodes against the solvent is
\[
\begin{split}
P_\mathrm{out} & = \sum_i \frac{\omega}{2 \pi}\int_{t_0}^{t_0 + 2\pi/\omega} \mathbf{f}_{p,i} \cdot \left( \frac{\partial \mathbf{p}_i^{\mathrm{NA}}}{\partial t}\right) dt \\
& = \sum_i \frac{\omega}{2 \pi} \int_{t_0}^{t_0 + 2\pi/\omega} \zeta \left\| \frac{\partial \mathbf{p}_i^{\mathrm{NA}}}{\partial t}\right\|^2 dt \\
& = \sum_i \frac{\omega}{2 \pi} \int_{t_0}^{t_0 + 2\pi/\omega} \zeta \left\|\mathbf{u}_i^\mathrm{NA}(\omega)\right\|^2\omega^2 \cos^2(\omega t + \theta^{\mathrm{NA}}) dt\\
& =   \frac{1}{2} \zeta \omega^2  \sum_i \left\|\mathbf{u}_i^{\mathrm{NA}}(\omega)\right\|^2 \\
& = \frac{1}{2} N \omega^2 \zeta \ell_0^2 \delta\gamma^2 \delta\Gamma(\omega)
\end{split}
\]
Since $P_\mathrm{in} = P_\mathrm{out}$, we have
\[
K''(\omega)-\eta_f\omega = \rho \omega \zeta \ell_0^2 \delta\Gamma(\omega)
\]
in which $\rho = N/V$, hence
\[\eta'(\omega) - \eta_f = \rho \zeta \ell_0^2 \delta\Gamma(\omega)
\]
For a quasistatic shear strain step $\delta\gamma$, the static nonaffinity, in terms of the individual static nonaffine displacements $\mathbf{u}_{i,\infty}^\mathrm{NA}$, is
\[
\delta\Gamma_\infty = \frac{1}{N\ell_0^2\delta\gamma^2}\sum_i \|\mathbf{u}_{i,\infty}^\mathrm{NA}\|^2
\]
Note that since the static nonaffine displacement vector must be the same as the frequency-dependent nonaffine displacement vector in the zero-frequency limit, i.e.\ $\mathbf{u}_{i,\infty}^\mathrm{NA} = \mathbf{u}_i^\mathrm{NA}(\omega\to 0)$, we have $\delta\Gamma_\infty = \delta\Gamma(\omega\to 0)$. Thus, we can write the zero-shear viscosity $\eta_0 = \eta'(\omega\to 0)$ in terms of the static nonaffinity as
\[
\eta_0 - \eta_f = \rho \zeta \ell_0^2 \delta\Gamma_\infty
\]

\section{Relaxation time}

We extract $\tau_c$ from the slope, on a log-linear plot, of the terminal exponential decay of $(\delta\sigma(t)-\delta\sigma_\infty)/\delta\sigma(0)$ vs $t$, as indicated in the inset for $\gamma_0<\gamma_c$. Specifically, we calculate the slope of the final $n=5$ points exceeding a sufficiently small threshold of $(\delta\sigma(t)-\delta\sigma_\infty)/\delta\sigma(0) = 10^{-6}$.

Another reasonable way of computing the relaxation time is (see Ref. \cite{saitoh_stress_2020})
\[
\tau_c = \lim_{\omega\to0}\frac{K'(\omega) - K_\infty}{\omega K''(\omega)} \equiv \lim_{\omega\to0}\frac{\left(K'(\omega)-K_\infty\right)/\omega^2}{\eta'(\omega)}
\]
which we can express in terms of $K(t)$. We can compute the dynamic moduli from the relaxation modulus as \cite{ferry_viscoelastic_1980}
\[
K'(\omega)  = K_\infty +  \omega \int_0^\infty \sin(\omega t) \left[K(t) - K_\infty \right] dt 
\]
and
\[
K''(\omega)  = \omega \int_0^\infty \cos(\omega t) \left[K(t)-K_\infty\right] dt
\]
Thus
\[
\lim_{\omega\to 0} \frac{K'(\omega) - K_\infty}{\omega^2} =  \int_0^\infty t (K(t) - K_\infty) dt
\]
and
\[
\lim_{\omega\to 0} K''(\omega)/\omega = \int_0^\infty  (K(t) - K_\infty) dt
\]
Plugging these in, we find
\[\tau_c =\frac{\int_0^\infty t (K(t) - K_\infty) dt}{\int_0^\infty  (K(t) - K_\infty) dt}=\frac{\int_0^\infty t (K(t) - K_\infty) dt}{\rho \zeta \ell_0^2 \delta\Gamma_\infty}
\]

\begin{figure}[htb!]
\centering
\includegraphics[width=0.4\columnwidth]{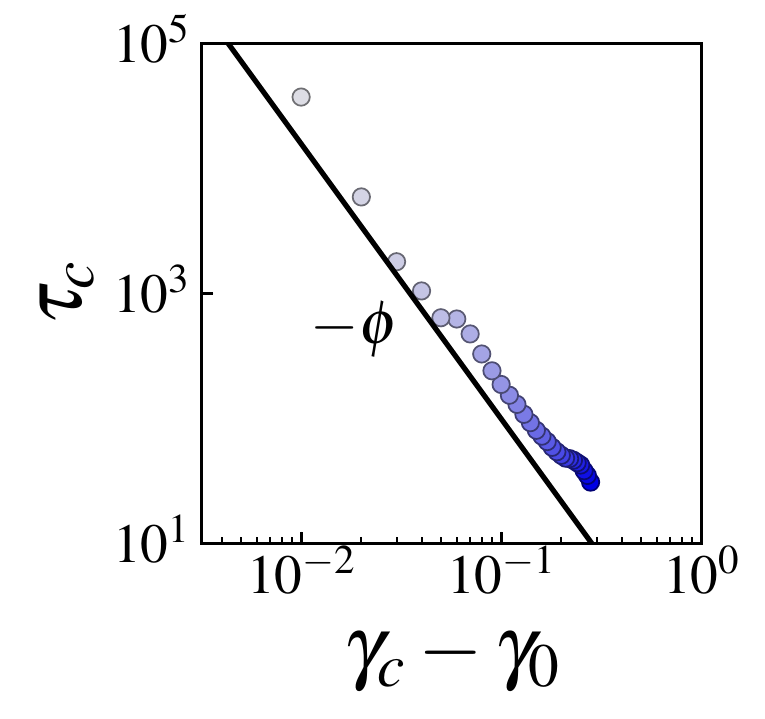}
\caption{\label{tau_scaling} Scaling of the slowest relaxation time $\tau_c$ with distance to the critical strain $\gamma_c - \gamma_0$, corresponding to the data presented in Fig. 2b in the main text. We observe excellent agreement with the predicted scaling $\tau_c \propto (\gamma_c - \gamma_0)^{-\phi}$ with $\phi = 2.2$.}
\end{figure}

In Fig. \ref{tau_scaling}, we plot $\tau_c$ vs. $\gamma_c - \gamma$ for a spring network with $N=6400$, $z=3.5$, $\tilde{\kappa}=0$ and $\delta\gamma = 10^{-4}$. The same data is plotted vs. $\gamma_0$ in Fig. 2b in the main text.

\section{Finite size effects}

We expect to observe the scaling relationships described in the main text when the correlation length is smaller than the system size. If $\tau_c$ diverges as $|\gamma_0-\gamma_c|^{-\phi}$ in the $L\to\infty$ limit, then we expect
\[
\tau_c\propto W^{\phi/\nu} \mathcal{A}(L^{1/\nu}(\gamma_0-\gamma_c))
\]
implying that we should observe $\tau_{c}(\gamma_c)\propto L^{\phi/\nu}$, a plot of $\tau_c L^{-\phi/\nu}$ vs. $L^{1/\nu}(\gamma_0-\gamma_c)$ for varying $L$ and $\gamma_0$ should yield a collapse.
For the static differential nonaffinity, prior work has shown \cite{shivers_scaling_2019} $\delta\Gamma\propto L^{\lambda/\nu} \mathcal{B}(L^{1/\nu} (\gamma_0-\gamma_c))$, so we expect the same finite-size scaling for the zero-shear viscosity,
\[
\eta_0-\eta_f \propto L^{\lambda/\nu} \mathcal{C}(L^{1/\nu} (\gamma_0-\gamma_c))
\]
such that $\eta_{0}(\gamma_c)-\eta_f \propto L^{\lambda/\nu}$. Likewise, we should see a collapse of $(\eta_0 - \eta_f) L^{-\lambda/\nu}$ vs. $L^{1/\nu}(\gamma_0-\gamma_c)$ for varying $L$ and $\gamma_0$. Here, $\mathcal{A}$, $\mathcal{B}$, and $\mathcal{C}$ are scaling functions.

\section{Nonaffinity and viscosity in soft sphere suspensions}

We will now briefly explore the response of dense suspensions of frictionless soft spheres in two and three dimensions near the onset of rigidity (jamming). As noted in the main text, prior work \cite{andreotti_shear_2012} has pointed out a connection between the zero-shear viscosity and quasistatic nonaffine velocity fluctuations in suspensions under steady shear.  To highlight the connection between nonaffinity and viscosity we discussed in the main text (Eq. 2), we will demonstrate here that the static differential nonaffinity is equivalent to, and diverges as $\phi_0\to\phi_j$ with the same exponent as, the excess viscosity.

These systems rigidify at a $d$-dimensional critical sphere volume fraction $\phi_j$, with $\phi_{j,2D} \approx 0.84$ and $\phi_{j,3D} \approx 0.64$. Under steady shear conditions, dense suspensions at volume fractions below $\phi_j$ these have been shown to exhibit a zero-shear viscosity that scales with the volume (or area) fraction as $\eta_0 \propto (\phi_j - \phi_0)^{-\beta}$, in which $\beta$ is an exponent generally reported in the range 2--2.8 in both simulations \cite{ikeda_disentangling_2013, kawasaki_diverging_2015,olsson_dimensionality_2019,ikeda_universal_2020,nishikawa_relaxation_2021,wang_constant_2015} and experiments \cite{boyer_unifying_2011,russel_divergence_2013}.

\begin{figure}[htb!]
\centering
\includegraphics[width=0.7\columnwidth]{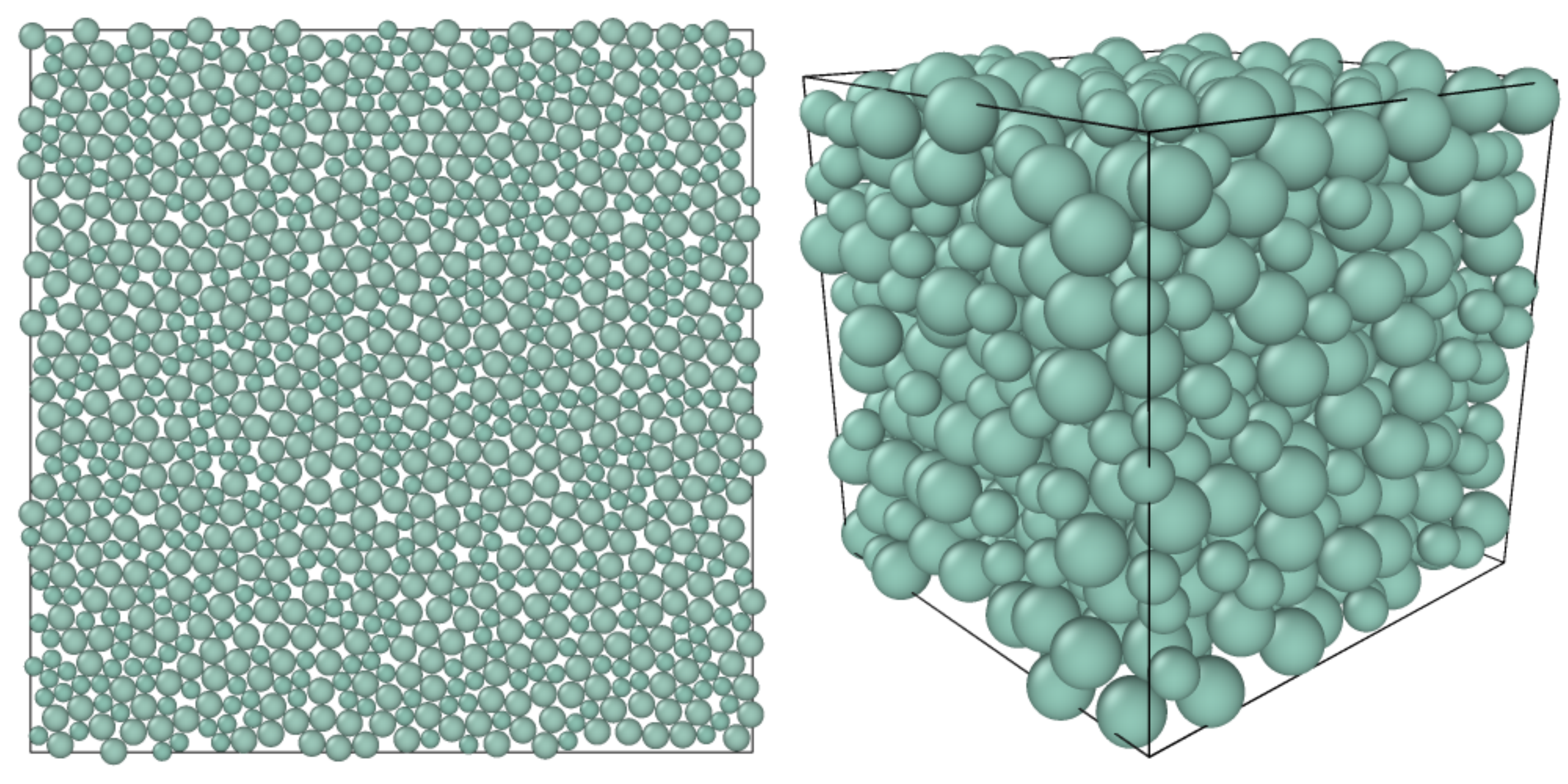}
\caption{ \label{fig_spheres_images} (a) Radially bidisperse assemblies of $N = 1000$ spheres in (left) $d=2$ with area fraction $\phi_0 = 0.84$ and (right) $d=3$ with volume fraction $\phi_0 = 0.64$. Images prepared using Ovito \cite{stukowski_visualization_2009}.}
\end{figure}

We consider $N$ spheres with diameters split evenly between $d_i\in(d_0, 1.4 d_0)$ to avoid crystallization \cite{koeze_mapping_2016}. The energy of a configuration with positions $\mathbf{r}_i$ is
\[
U = \frac{\mu}{2}\sum_{i}\sum_{j>i} \left( 1-\|\mathbf{r}_j - \mathbf{r}_i\|/d_{ij}\right)^2 \Theta \left( 1-\|\mathbf{r}_j - \mathbf{r}_i\|/d_{ij}\right)
\]
in which $d_{ij} = (d_i + d_j)/2$ and $\Phi$ is the Heaviside step function. To prepare initial configurations, we first randomly place the spheres in a $d$-dimensional box of side length $3L$ and quasistatically compress the system in small steps to a final side length $L$, chosen to yield the specified sphere volume fraction $\phi_0$. Then, to produce a configuration consistent with slowly applied steady shear strain, we quasistatically apply (again in small steps) an initial simple shear of $\gamma_0 = 5$. Sample configurations are shown in Fig. \ref{fig_spheres_images}.

\begin{figure}[htb!]
\centering
\includegraphics[width=0.8\columnwidth]{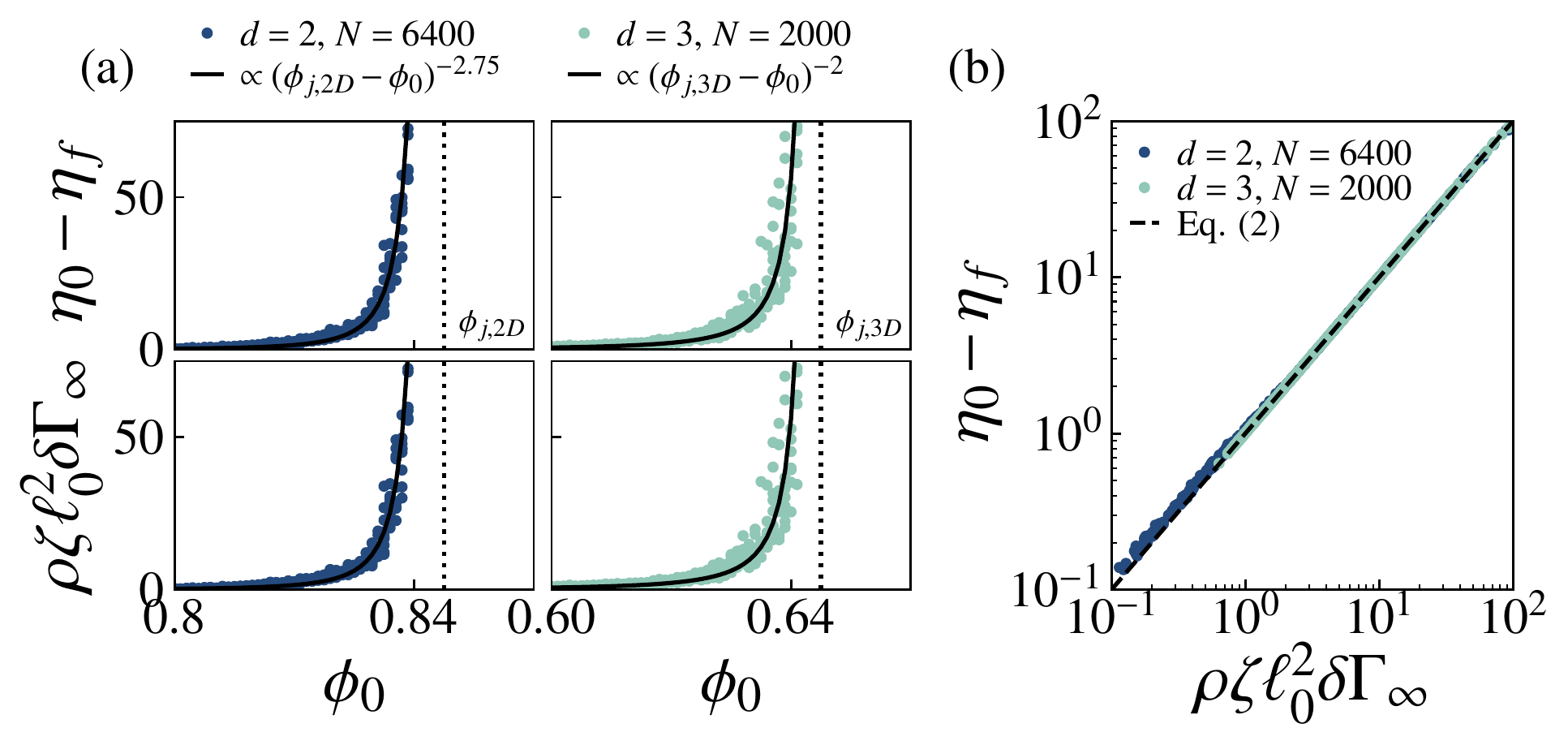}
\caption{ \label{fig_spheres_data} (a) Excess zero-shear viscosity $\eta_0 - \eta_f$ and scaled static differential nonaffinity $\delta\Gamma_{\infty}$ for $N$ spheres in two and three dimensions as a function of volume (or area) fraction $\phi_0$, plotted with fits (solid lines) proportional to $(\phi_j - \phi_0)^{-\beta}$ as indicated above each column. Each point represents a measurement for a randomly generated sample, with $10$ samples for each $\phi_0$. Here, we use $\phi_{j,2D} = 0.845$ and $\phi_{j,3D} = 0.645$. (b) Eq. 2 (main text) is satisfied for both $d=2$ and $d=3$. Data are the same as in (a). }
\end{figure}

Using the pre-sheared initial configuration, we follow the same stress relaxation procedure described earlier for networks, with $\delta\gamma = 10^{-5}$, and compute both the excess zero-shear viscosity $\eta_0-\eta_f$ and static differential nonaffinity $\delta\Gamma_\infty$ as a function of volume fraction. For both $d=2$ and $d=3$, we observe values of the scaling exponent $\beta$ consistent with the range reported in the literature (see Fig. \ref{fig_spheres_data}a) and find that the relationship between zero-shear viscosity and static differential nonaffinity provided in Eq. 2 (main text), i.e.\ $\eta_0-\eta_f = \rho \zeta \ell_0^2 \delta\Gamma_\infty$, is consistently satisfied (see Fig. \ref{fig_spheres_data}b).  These results suggest that the static differential nonaffinity, which can be inexpensively computed by energy minimization after a single small shear strain step, could provide a complementary route for concretely determining the viscosity divergence exponent $\beta$ in this and other related systems (e.g.\ suspensions of frictional spheres \cite{singh_shear_2020}). Note that a different viscosity divergence exponent $\beta \approx 1.5$ is observed for initial configurations generated without pre-shear \cite{saitoh_stress_2020}. Because the static differential nonaffinity necessarily scales with $\beta$ irrespective of preparation, our results suggest that the distinction in scaling between isotropically compressed and pre-sheared suspensions is due to differences in their nonaffine response.

\newpage

\section{Definition of $\gamma_c$ and expanded collapse plots}

\begin{figure}[htb!]
\centering
\includegraphics[width=0.9\columnwidth]{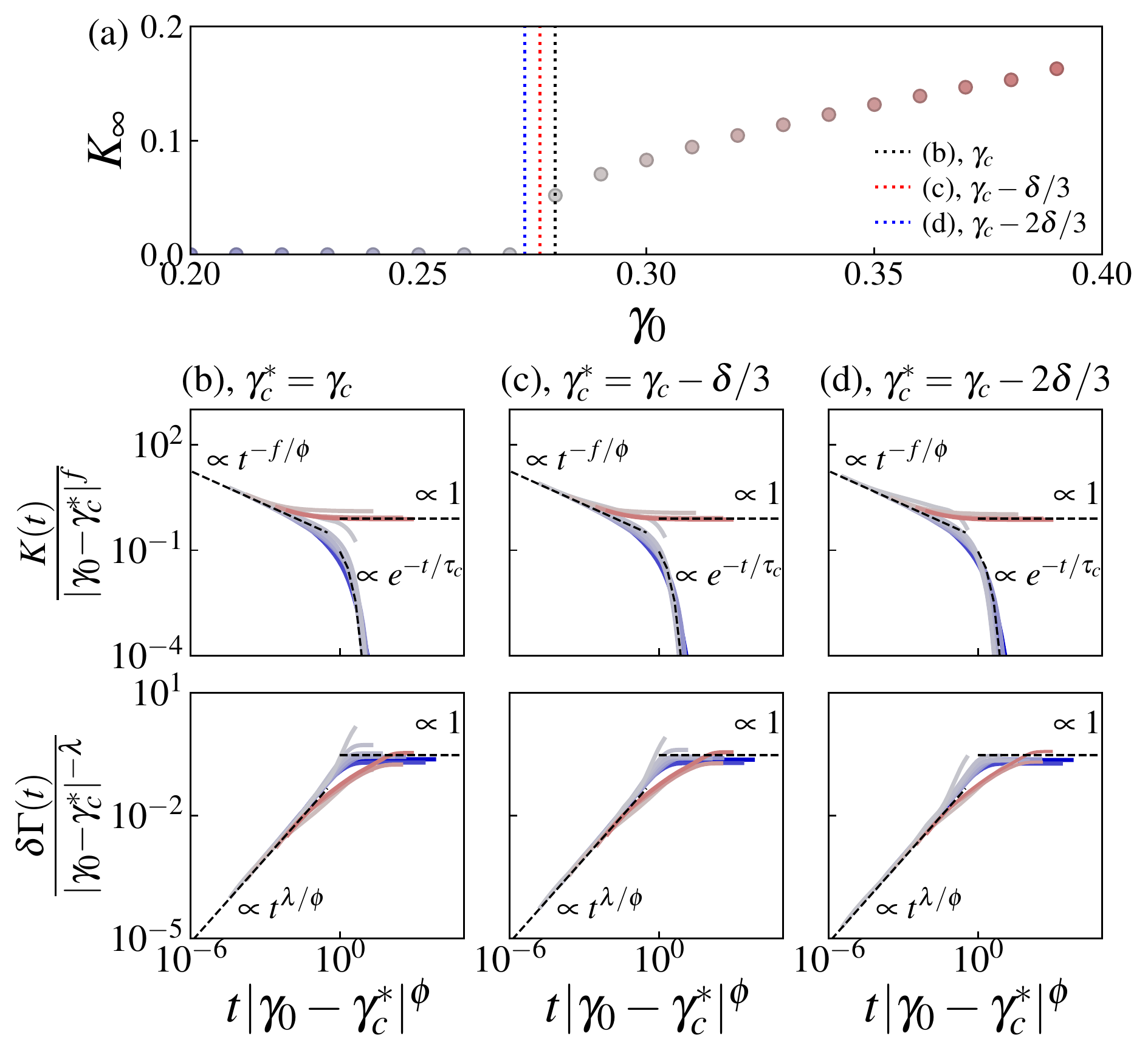}
\caption{ \label{supp_fig_collapse} (a) Quasistatic differential shear modulus $K_{\infty}$ vs. prestrain $\gamma_0$, reproduced from Fig. 2b in the main text. The dotted lines represent possible values of the critical strain $\gamma_c$ given the spacing $\delta$ between consecutive values of $\gamma_0$. The black dotted line represents the value of $\gamma_c$ used in the main text, i.e. the first value of $\gamma_0$ at which the measured $K_{\infty}$ is nonzero. Because the true value of $\gamma_c$ is between $\gamma_c - \delta$ and $\gamma_c$, we can also consider other values in this range: the red and blue dotted lines correspond $\gamma_c - \delta/3$ and $\gamma_c - 2\delta/3$, respectively. (b-c) Reproducing the collapse plots of Fig. 2c (main text) using these values of $\gamma_c^*$, we observe little variation in the quality of the collapse, suggesting that the spacing $\delta$ is sufficiently small. Note that (b) is simply a reproduction of Fig. 2c in the main text.}
\end{figure}

For networks with $\tilde{\kappa} = 0$, we define the critical strain $\gamma_c$ as the lowest value of the applied prestrain $\gamma_0$ at which the quasistatic differential shear modulus $K_{\infty}$ is measurably nonzero, exceeding a cutoff of $10^{-6}$. The precision of our determination of $\gamma_c$ is thus limited by the spacing between $\gamma_0$ points, which we designate $\delta$. The \textit{true} value of the critical strain lies somewhere between our measured value $\gamma_c$ and the previous strain point $\gamma_c - \delta$. Fig. \ref{supp_fig_collapse}a shows a zoomed-in view of the region of the plot of $K_{\infty}$ vs. $\gamma_0$ provided in Fig. 2c in the main text. 

To verify that our chosen spacing $\delta$ is sufficiently small, we check the dependence of the quality of the collapse plots in Fig. 2c (main text) on the chosen critical strain value. Let $\gamma_c$ denote the critical strain defined in the main text, i.e. the first strain at which $K_{\infty}$ is nonzero, and $\gamma_c^*$ the critical strain value to be used in the collapse plots in Fig. \ref{supp_fig_collapse}. The black dotted line in Fig. \ref{supp_fig_collapse} indicates the value used in the main text, $\gamma_c^* = \gamma_c$, while the red and blue dotted lines represent $\gamma_c^* = \gamma_c - \delta/3$ and $\gamma_c^* = \gamma_c - 2\delta/3$, respectively.
 Figs. \ref{supp_fig_collapse}b-d show the corresponding collapse plots. Note that Fig. \ref{supp_fig_collapse}b is simply an expanded version of Fig. 2c. We find that the quality of the collapse is insensitive to the choice of $\gamma_c^*$, suggesting that our chosen spacing $\delta$ is indeed sufficiently small.

\newpage

\bibliographystyle{apsrev}
\bibliography{Bibliography}{}